\documentclass[12pt,twoside]{article}
\usepackage{inputenc}
\usepackage[english,russian]{babel} 
\usepackage{vestnik_eng}
\usepackage{amsfonts}
\usepackage{amsmath}
\usepackage{amssymb}
\usepackage{amsthm}
\usepackage{graphicx}          
\usepackage[usenames]{color}
\usepackage{hyperref}
\begin{document}
\def\No{\textnumero}
\def\Re{\mathop{\rm Re}\,}
\def\Im{\mathop{\rm Im}\,}
\def\dom{\mathop{\rm dom}\,}
\def\dist{\mathop{\rm dist}}
\def\grad{\mathop{\rm grad}}
\renewcommand{\proof}{\vspace{2mm}\hspace{-7mm}\textit{Proof.}}
\renewcommand{\endproof}{\begin{flushright} \vspace{-2mm}$\Box$\vspace{-4mm}
\end{flushright}}
\newcommand{\phan}{\hspace*{0cm}}
\newcommand{\comment}{}

\noindent
{\textit{
Khrapov S.S., Khoperskov S.A., Khoperskov A.V.  New features of parallel implementation of N-body problems on GPU  // Bulletin of the South Ural State University, Series: Mathematical Modelling, Programming and Computer Software, 2018, v.11, no.1, p.124--136.} \\
\noindent DOI: 10.14529/mmp180111 \quad http://mmp.vestnik.susu.ru/article/en/495
}

\hrule

\begin{flushleft}
\textbf{MSC 34N05, 37M05, 68U20} 
\end{flushleft}
\author{S.S. Khrapov, {\rm  Volgograd State University, Volgograd, Russian Federation, khrapov@volsu.ru}, \\ S.A. Khoperskov,  {\rm   Institute of Astronomy, Russian Academy of Sciences, Moscow, Russian Federation \\ sergey.khoperskov@gmail.com},  \\ A.V. Khoperskov, {\rm  Volgograd State University, Volgograd, Russian Federation, khoperskov@volsu.ru}}
\title{NEW FEATURES OF PARALLEL IMPLEMENTATION OF N-BODY PROBLEMS ON GPU}
\maketitle{NEW FEATURES OF PARALLEL IMPLEMENTATION OF N-BODY PROBLEMS ON GPU}
\begin{abstract} \begin{tabular}{p{0mm}p{139mm}}
&\noindent {\footnotesize \qquad

This paper focuses on the parallel implementation of a direct $N$-body method~(particle-particle algorithm) and the application of multiple GPUs for galactic dynamics simulations. Application of a hybrid OpenMP-CUDA technology is considered for models with a number of particles $N \sim 10^5 \div 10^7$. By means of $N$-body simulations of gravitationally unstable stellar galactic we have investigated the algorithms parallelization efficiency for various Nvidia Tesla graphics processors~(K20, K40, K80). Particular attention was paid to the parallel performance of simulations and accuracy of the numerical solution by comparing single and double floating-point precisions~(SP and DP). We showed that the double-precision simulations are slower by a factor of~$1.7$ than the single-precision runs performed on Nvidia Tesla K-Series processors. We also claim that application of the single-precision operations leads to incorrect result in the evolution of the non-axisymmetric gravitating $N$-body systems. In particular, it leads to significant quantitative and even qualitative distortions in the galactic disk evolution.  For instance, after $10^4$ integration time steps for the single-precision numbers the total energy, momentum, and angular momentum of a system with $N = 2^{20}$ conserve with accuracy of $10^{-3}$, $10^{-2}$ and $10^{-3}$ respectively, in comparison to the double-precision simulations these values are $10^{-5}$, $10^{-15}$ and $10^{-13}$, respectively.  Our estimations evidence in favour of usage of the second-order accuracy schemes with double-precision numbers since it is more efficient than in the fourth-order schemes with single-precision numbers.

\qquad\keywordsanglish{Multi-GPU, OpenMP-CUDA, GPU-Direct, NVIDIA TESLA, N-body, single and double precision numerical simulation, collisionless system, gravitational instability.}}
\end{tabular}\end{abstract}

\markboth{S.S. Khrapov et al.}{Parallel implementation features of numerical N-body models on GPU}


\section*{Introduction}
\hspace{0.7 cm}Different $N$-body models are essential for the theoretical studies of the gravitating collisionless systems dynamics~\cite{bib1}, such as galactic stellar disks, elliptical galaxies, globular clusters, galactic dark matter haloes~\cite{KennedyEtal2016,bib11,SmirnovSotnikova2017}. $N$-body models is a fundamental tool for cosmological dark matter only simulations \cite{Klypin2017,KnebeEtal2018}.

Our research may also contribute to the Lagrangian methods of computational fluid dynamics. Let us refer to a widely used SPH method (Smooth Particle Hydrodynamics) for a self-gravitating gas~\cite{bib2,PortaluriDebattista2017}.
In addition to astrophysical applications, the $N$-body method is widely used for the modeling of rarefied plasma, ion and electron beams, problems of molecular dynamics, which differ in the type of interaction between the particles.

Gravitational interactions between $N$ particles is a resource-intensive problem, and it can be solved using different approaches. There are various groups of approximate methods (for example, Particle-Mesh, SuperBox \cite{bib10}, Particle-Multiple-Mesh/Nested-Grid Particle-Mesh~\cite{bib3}, TreeCode~\cite{bib4}, Fast Multiple Method~\cite{bib5}) which significantly reduce computation time in comparison to the direct calculation of the gravitational forces between all pairs of particles (so-called Particle-Particle or PP method) which has a complexity of~$O(N^2)$. However, the PP approach provides the best accuracy for the total gravitational force calculation, and it is a kind of benchmark for the testing of the approximate methods.

In computational astrophysics, the problem of the software transfer to new hardware platforms becomes relevant due to the wide distribution of powerful computer systems on graphics processors. The result of the parallel software implementation and its efficiency depend a lot on the features of the code and the sequence of numerical operations~\cite{Huang2016,Steinberg2017}.

 To increase the spatial resolution for a large number of particles $N$ researchers often use a second-order numerical time integration schemes with a single-precision numbers. From another hand, such schemes are very efficient for long-term integration, e.g., evolution of the galactic systems at cosmological time scales (up to hundred of disk rotations or $\sim 10^5-10^6$ integration time steps).

This approach is justified for CPU calculations, however for parallel GPUs, this approach can lead to unphysical results because of parallel features of the hardware. Our work aims to provide computational characteristics analysis of the parallel program for the self-gravitating system simulation by means of a direct $N$-body method implemented on GPUs.

\section{Basic Equations and Numerical Scheme}
\hspace{0.7 cm}Equations of motion for an $N$-body model can be written as following:
\begin{equation}
\label{eq:motionEq}
\frac{d \mathbf{v}_i}{d t} = \sum_{j=1,j\neq i}^N{\mathbf{f}_{ij}}\,, \qquad i=1,...,N\,,
\end{equation}
where $\mathbf{v}_i$ is the velocity vector of the $i$-th particle.
The gravitational interaction force between $i$-th and $j$-th particles is:
\begin{equation}
\label{eq:gravity}
\mathbf{f}_{ij} = -G  {\frac{m_j \,(\mathbf{r}_i - \mathbf{r}_j)}{|\mathbf{r}_i - \mathbf{r}_j + \delta|^3}}\,,
\end{equation}
where $G$ is the gravitational constant, $m_j$ is the mass of $j$-th particle, $\delta$ is the gravitational softening length, the radius-vector $\mathbf{r}_i(t)=\int \mathbf{v}_i\, dt$ determines the position of the $i$-th particle in space and for cartesian coordinate system we can use the following $|\mathbf{r}_i - \mathbf{r}_j + \delta|=\sqrt{(x_i-x_j)^2 + (y_i-y_j)^2+(z_i-z_j)^2 + \delta^2}$. The small parameter $\delta$ ensures a collisionlessness of the system.

\begin{figure}[!h]
	\centering
	\includegraphics[width=0.99\hsize]{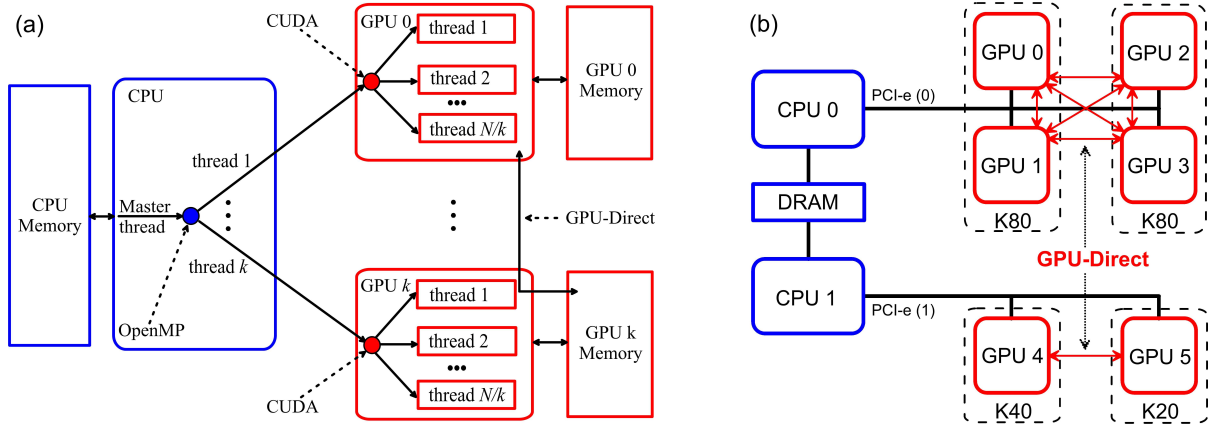}
	\vskip 4 mm
	\caption{(a) The two-level scheme of parallelization with OpenMP--CUDA. (b) Architecture of 2$\times$CPU+6$\times$GPU.}
\end{figure}

Equilibrium model of the collisionless stellar disk in the radial direction is made by the balance between the disk self-gravity, rotation and chaotic (thermal) motions \cite{bib7}:
\begin{equation}
\label{eq:equlib}
\frac{v_\varphi}{r}= -\frac{\partial\Psi}{\partial r}+\frac{c_r^2}{r}\left(1-\frac{c_\varphi^2}{c_r^2}+\frac{r}{\varrho c_r^2}
\frac{\partial (\varrho c_r^2)}{\partial r}+ \frac{r}{c_r^2}\frac{\partial \langle{v_r v_z}\rangle}{\partial z}\right)\,,
\end{equation}
here $\Psi$ is the gravitational potential, $\varrho$ is the density of the disk, $(r, \varphi, z)$ is the cylindrical coordinates, $c_r$ is the radial velocity dispersion, $c_\varphi$ is the azimhutal velocity dispersion, $v_r, v_\varphi, v_z$ are the radial, azimuthal and vertical velocity components, respectively, $\langle ... \rangle$ is the averaging operation.

Vertical equilibrium for the geometrically thin disk is the following:
\begin{equation}
\label{eq:equlib_z}
\varrho\frac{d^2 \varrho}{dz^2} - \left[ \frac{d\varrho}{dz} \right]^2 + \frac{4\pi G\varrho^2}{c_z^2} \left[ \varrho - \varrho_* \right] + \frac{\varrho^2}{c_z^2} \frac{d}{dz}\frac{\varrho_\alpha}{\varrho} = 0
\,,
\end{equation}
where $\displaystyle \varrho_* = \frac{1}{4{\pi}Gr}\frac{\partial V_c^2}{\partial r}$, $\displaystyle \varrho_\alpha = \frac{\partial}{r\,\partial r}\left( r\varrho\langle v_r v_z  \rangle \right)$, $V_c$ is the circular velocity in the disk equatorial plane~($ z = 0 $). In Eqs~(\ref{eq:motionEq}), (\ref{eq:gravity}) we neglect the gravitational interaction between dark matter halo, stellar bulge and gaseous component~\cite{bib10,bib8,bib9}. According to this model iterative procedure of the disk initial conditions generation is described in Ref.~\cite{bib1,bib7,RodionovAthanassoula2009}. Note that, in such model, the stellar disk is completely self-gravitating, and the following analysis provides an upper limit for numerical errors in direct $N$-body integration.

\begin{figure}[!h]
	\centering
	\includegraphics[width=0.99\hsize]{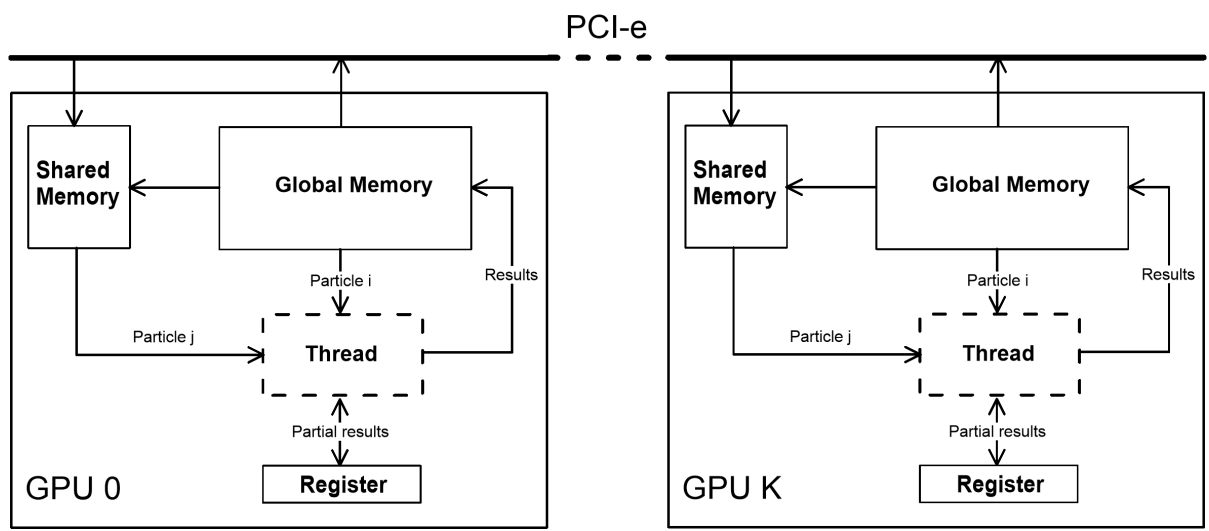}
	\vskip 4 mm
	\caption{Scheme of the parallel algorithm for calculation of the gravitational interactions between particles on Multi-GPU}
	\label{fig:CUDAMemory}
\end{figure}
Next, we describe the main features of time integration of $N$-body system. For the second-order time integration of equations ~(\ref{eq:motionEq}), we used a leapfrog scheme with a fixed step size of $0.2$~Myr. Such approach is also called as ``Kick-Drift-Kick"\, or KDK scheme. In order to speed up the integration this method utilizes a gravity solver for Eq.~(\ref{eq:motionEq}) only once at each time step. The main steps of the leapfrog method for self-gravity N-body models are as follows:

{(I) Velocity vector $\mathbf{v}_i$ at a predictor sub-step, at a moment of $t + \Delta t$, is given by}
\begin{equation}
\label{eq:predictor}
\widetilde{\mathbf{v}}_i(t+\Delta t) = \mathbf{v}_i(t) + \Delta t \, \sum_{j=1,\,j\neq i}^N{\mathbf{f}_{ij}(t)}\,,
\end{equation}
where $\Delta t$ is a full time step.

{(II) Next we update the position of particles $\mathbf{r}_i$ at time $t+\Delta t$ as}
\begin{equation}
\label{eq:rnew}
\mathbf{r}_i(t+\Delta t) = \mathbf{r}_i(t) + \frac{\Delta t}{2} \, \left[\widetilde{\mathbf{v}}_i(t+\Delta t) + \mathbf{v}_i(t)\right]\,.
\end{equation}
After this step we need to re-calculate the accelerations of the particles $\mathbf{f}_{ij}(t + \Delta t)$ according to Eq.~(\ref{eq:gravity}).

{(III) During the corrector step the velocity, $\mathbf{v}_i$, values are recalculated at time $t+\Delta t$ as:}
\begin{equation}
\label{eq:corrector}
\mathbf{v}_i(t+\Delta t) = \frac{\mathbf{v}_i(t)+\widetilde{\mathbf{v}}_i(t+\Delta t)}{2} + \frac{\Delta t}{2} \, \sum_{j=1,\,j\neq i}^N{\mathbf{f}_{ij}(t+\Delta t)}\,.
\end{equation}

As it is clearly seen, the KDK scheme (\ref{eq:predictor})~--~(\ref{eq:corrector}) allows to increase an $N$-body solver performance by a factor of $2$ in comparison to Runge-Kutta schemes of the second order approximation. This is because the gravitational interaction between the particles is calculated only once per integration time step.


\begin{figure}[!h]
	\centering
	\includegraphics[width=0.99\hsize]{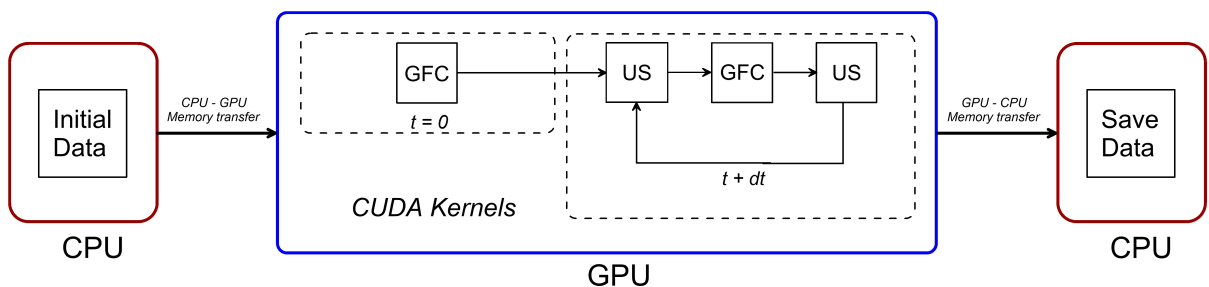}
	\vskip 0 mm
	\caption{Flow diagram for the calculation module.}
\end{figure}

\section{Parallel Algorithm Structure}
\hspace{0.7 cm}In this section, we describe in detail the parallel algorithm of the $N$-body problem integration based on Hybrid Parallelization Technology OpenMP-CUDA~(Fig.~1,~2).

To use multiple GPUs for $N$-body system integration, we parallelized the algorithm described above (\ref{eq:predictor})~--~(\ref{eq:corrector}) by using OpenMP-CUDA and GPU-Direct technologies. Figures 1a and 1b demonstrate the schemes of the two-level parallelization of OpenMP-CUDA and the communications between GPUs based on GPU-Direct technology.

\begin{figure}[!h]
	\centering
	\includegraphics[width=0.9\hsize]{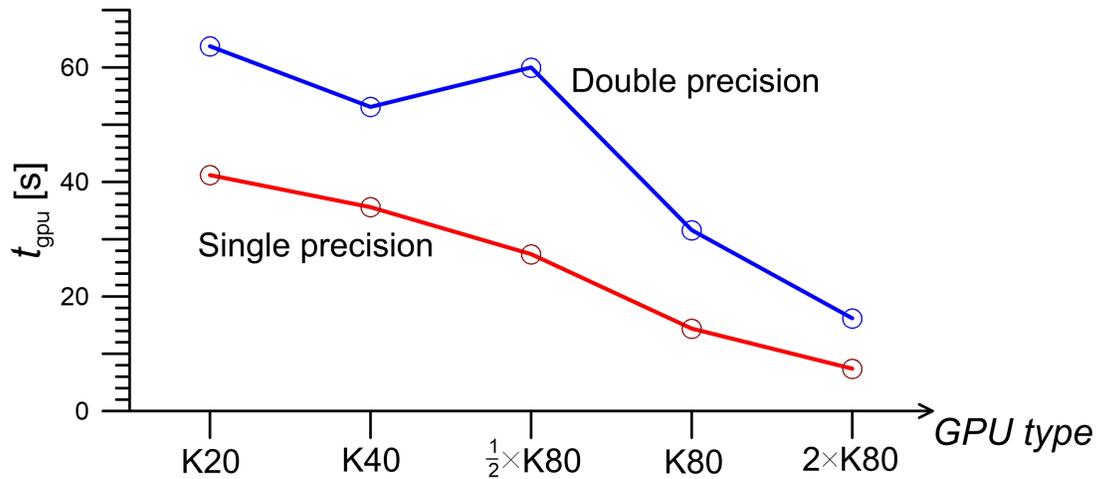}
	\vskip 0 mm
	\caption{Calculation time (in seconds) for galactic stellar disk evolution on various GPUs}
\end{figure}

According to hardware architecture restrictions, our parallel algorithm for $N$-body dynamics can be used only on computational systems with shared CPU + $k\,\times\,$GPU memory, taking into account two-level OpenMP-CUDA scheme (Fig.~1a). OpenMP technology allows to create $k$-parallel threads on the central processor (CPU) which runs tasks on $k \times$GPUs by using CUDA technology. In this case, each GPU calculates the dynamics of $N / k$ particles, and fast data exchange between the GPU is carried out via the PCI-e bus based on GPU-Direct technology. Note that for multiprocessor computing systems (2 or more CPUs on the same motherboard), GPU-Direct technology works only for GPUs connected to the PCI-e buses under the control of one processor (Fig.~1b). Previously we applied a similar approach for the hydrodynamic SPH-code \cite{bib11}.

Figure~2 shows the data flows between different types of GPU memory while the calculation of gravitational forces between particles (\ref{eq:gravity}) on Multi-GPU architectures. Each computational CUDA kernel first copies the data for the $i$ and $j$ particles from the slow Global Memory to the fast Shared Memory, and then, after the synchronization of all parallel CUDA threads, the gravitational interaction between the particles is calculated according to the Eq.~(\ref{eq:gravity}). This parallel algorithm allows us to accelerate the computations by a factor of~$3-4$ due to the fast Shared Memory \cite{bib11,bibl12}. We emphasize that the transfer of data, stored on different GPUs, from Global Memory to Shared Memory proceeds via PCI-e bus using GPU-Direct technology. The computer algorithm for calculation of the $N$-body system dynamics consists of two main Global CUDA Kernels, which are launched on a CPU with multiple GPUs by using OpenMP technology:

\noindent
--- The Gravity Force Computation (GFC) is a CUDA Kernel for calculation of the gravitational forces between particles (\ref{eq:gravity}). It is characterized by a computational complexity of $O(N^2)$.

\noindent
--- The Update System (US) is a CUDA Kernel for calculation of the particles positions and velocities at a next step of the KDK scheme (\ref{eq:predictor})---(\ref{eq:corrector}). Here computational complexity is $O(N)$.

Figure~3 shows the sequence of execution of the main Global CUDA Kernels.

\section{Main results}
\hspace{0.7 cm}We have studied the parallelization efficiency and accuracy of our algorithm by means of simulation of the gravitationally unstable collisionless disk~\cite{bib1,bib6,romeo2013}. The calculations were carried out on GPU Nvidia Tesla computers: K20, K40, K80.

In Figures~4 we show the computation time~(one integration step) for various GPUs. Calculation time with double precision on one Tesla K80 GPU is by 15\% larger than the Tesla K40 GPU. This is due to the different speeds of access to Global Memory and Shared Memory on these GPUs \cite{bib11}. The speed of access to global memory on the K80 GPU is greater than on the K40 GPU, and in the case of shared memory, the situation is the opposite.

In Table~1 we present the calculation time for simulations with various numbers of particles and performed on different GPUs for single- and double precision. Computation time depends quadratically on the number of particles, which corresponds to the $O(N^2)$ algorithm complexity. Table~1 also shows that increase of the number of GPUs leads to an almost linear increase in computing performance.

\begin{figure}[th]
	\begin{flushright}
		\bf{Table 1}\vspace{-2mm}
	\end{flushright}
\centerline{The dependence of the one integration step calculation time}
\centerline{on the number of particles $N$ obtained on GPU NVIDIA TESLA K80}
	\begin{center}
	\begin{tabular}{|c|c|c|c|c|c|c|}
	\hline
	& \multicolumn{3}{|c|}{$t_{gpu}$ [s], single precision} & \multicolumn{3}{|c|}{$t_{gpu}$ [s], double precision}\\ \hline
	$N\times1024$ & 1$\times$GPU & 2$\times$GPU & 4$\times$GPU & 1$\times$GPU & 2$\times$GPU & 4$\times$GPU\\ \hline
	128           & 0.4	         & 0.2          & 0.1          & 0.9          & 0.5          & 0.3\\ \hline
	256	          & 1.7          & 0.9          & 0.45         & 3.7          & 2            & 1 \\ \hline
	512	          & 6.9          & 3.6          & 1.8          & 15           & 7.9          & 4 \\ \hline
	1024          & 27.4         & 14.4         & 7.4          & 60           & 31.6         & 16.2 \\ \hline
	2048          & 109.6        & 57.6       	& 29.6         & 240          & 126.4        & 64.8 \\ \hline
	4096          & 438          & 230          & 118          & 960          & 506          & 259 \\ \hline
	8192          & 1754         & 922          & 474          & 3840         & 2022         & 1037 \\ \hline
\end{tabular}
	\end{center}
\end{figure}

\begin{figure}[!h]
	\centering \includegraphics[width=0.7\hsize]{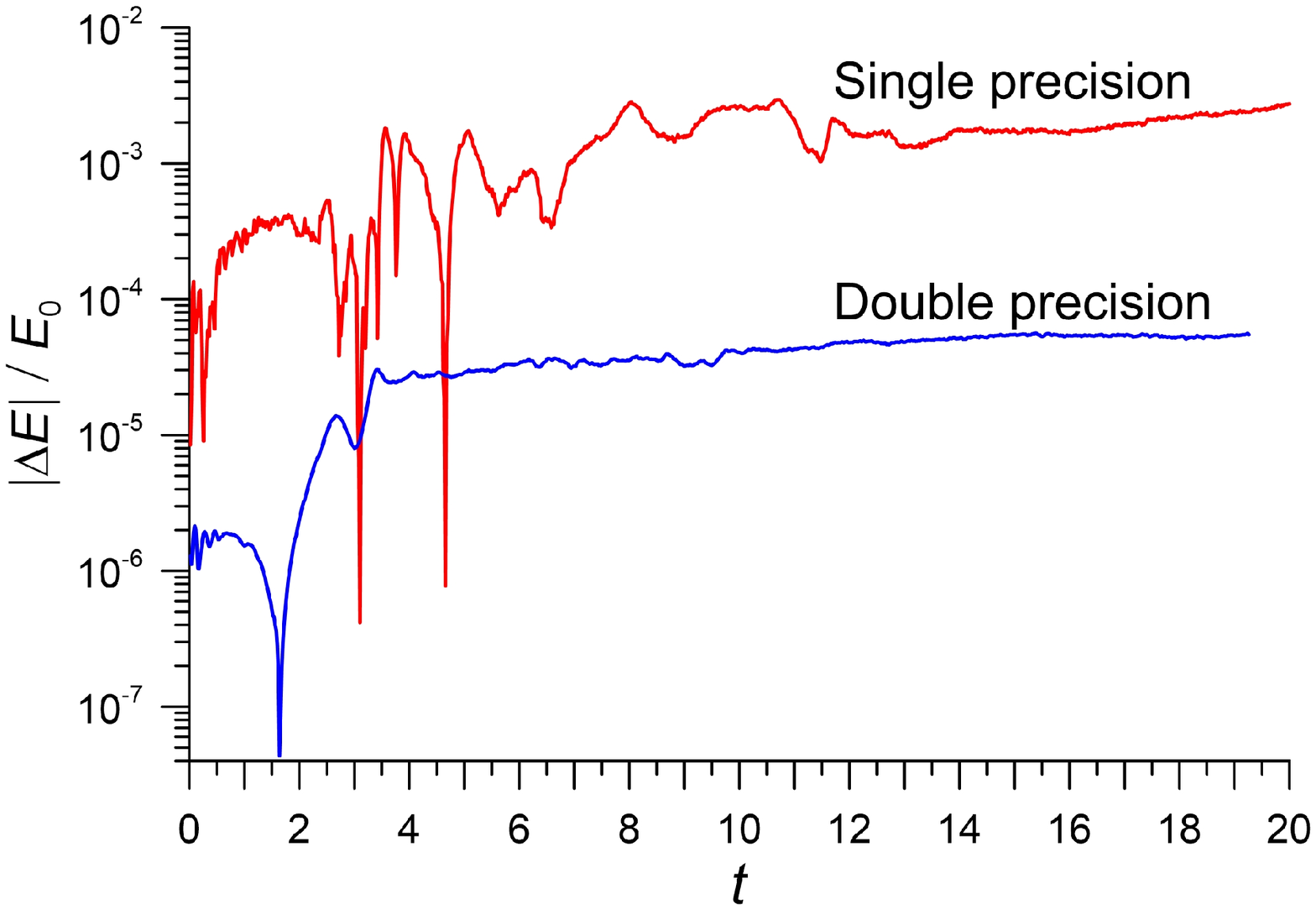}
	\centering \includegraphics[width=0.7\hsize]{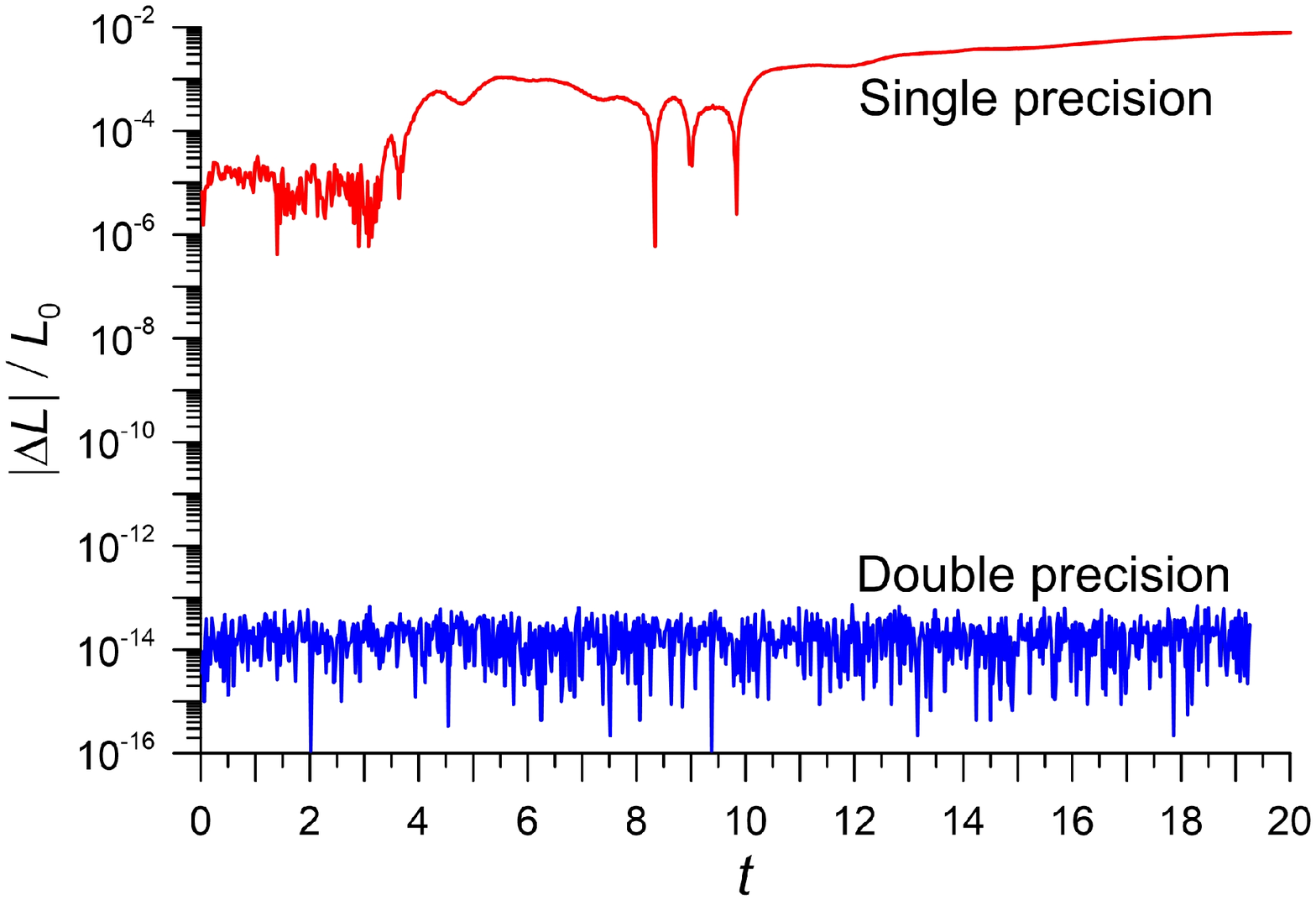}
	\vskip 0 mm
	\caption{Evolution of the relative deviations of the total energy (top panel) and angular momentum~(bottom panel) for the stellar disk system of $N = 2^{20}$ particles. Red and blue lines are the single precision and double precision, respectively. Period of rotation of the stellar disk at the periphery is $\approx 4$.}
\end{figure}

The speed of our algorithm on the Nvidia Tesla K-Series processor for double precision is smaller by a factor of $1.5-2.2$ in comparison to the single precision depending on the particles number and GPU type. The average speed up for a single GPU is approximately~$1.7$.

Let us consider the accuracy of the most important integral physical conservation laws.
 For total energy, momentum and angular momentum, we have the following expressions:
\begin{equation}\label{eq:energy}
   E = \sum_{i=1}^{N} \frac{m_i |\mathbf{v}_i|^2}{2} + \frac{1}{2} \sum_{i=1}^{N} \sum_{j=1,\, j\neq i}^{N} \frac{G m_i m_j}{|\mathbf{r}_i - \mathbf{r}_j + \delta|}
    \,,
\end{equation}
\begin{equation}\label{eq:momentum}
   \mathbf{P} = \sum_{i=1}^{N} m_i \mathbf{v}_i\,,
\end{equation}
\begin{equation}\label{eq:anglmomentum}
   L = \sum_{i=1}^{N} m_i [\mathbf{r}_i \times \mathbf{v}_i]_z\,.
\end{equation}

Figure~5 demonstrates the evolution of the total energy error for the entire system of particles. Obviously, this quantity accumulates faster for single precision. The angular momentum error for double precision varies in the range of $10^{-16} \div 10^{-13}$, while for single precision format it is larger by four orders of magnitude. Note, however, that center mass of the stellar disk system for SP and DP (Fig. 6) is roughly the same.

\begin{figure}[!h]
	\centering \includegraphics[width=0.7\hsize]{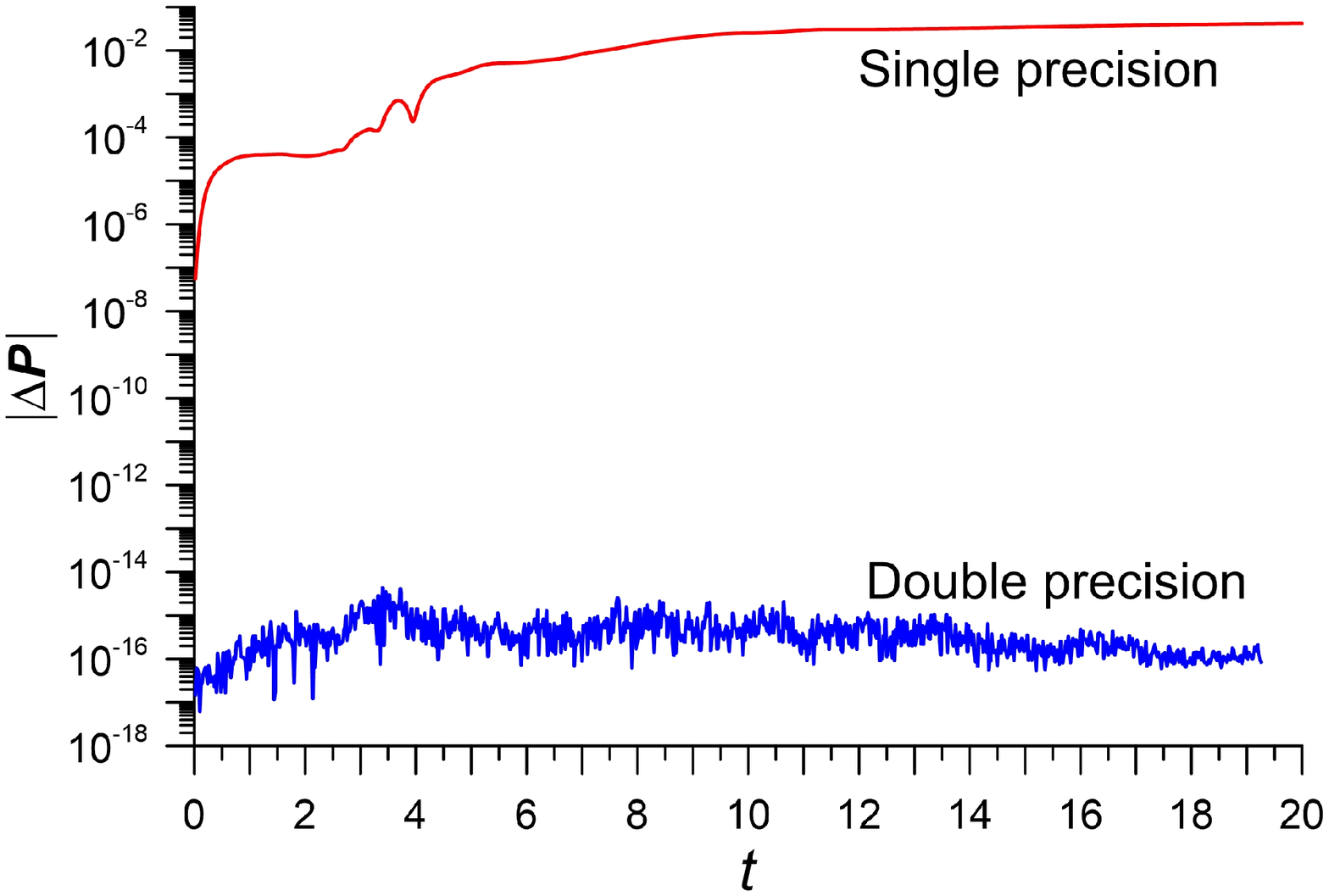}
	\centering \includegraphics[width=0.7\hsize]{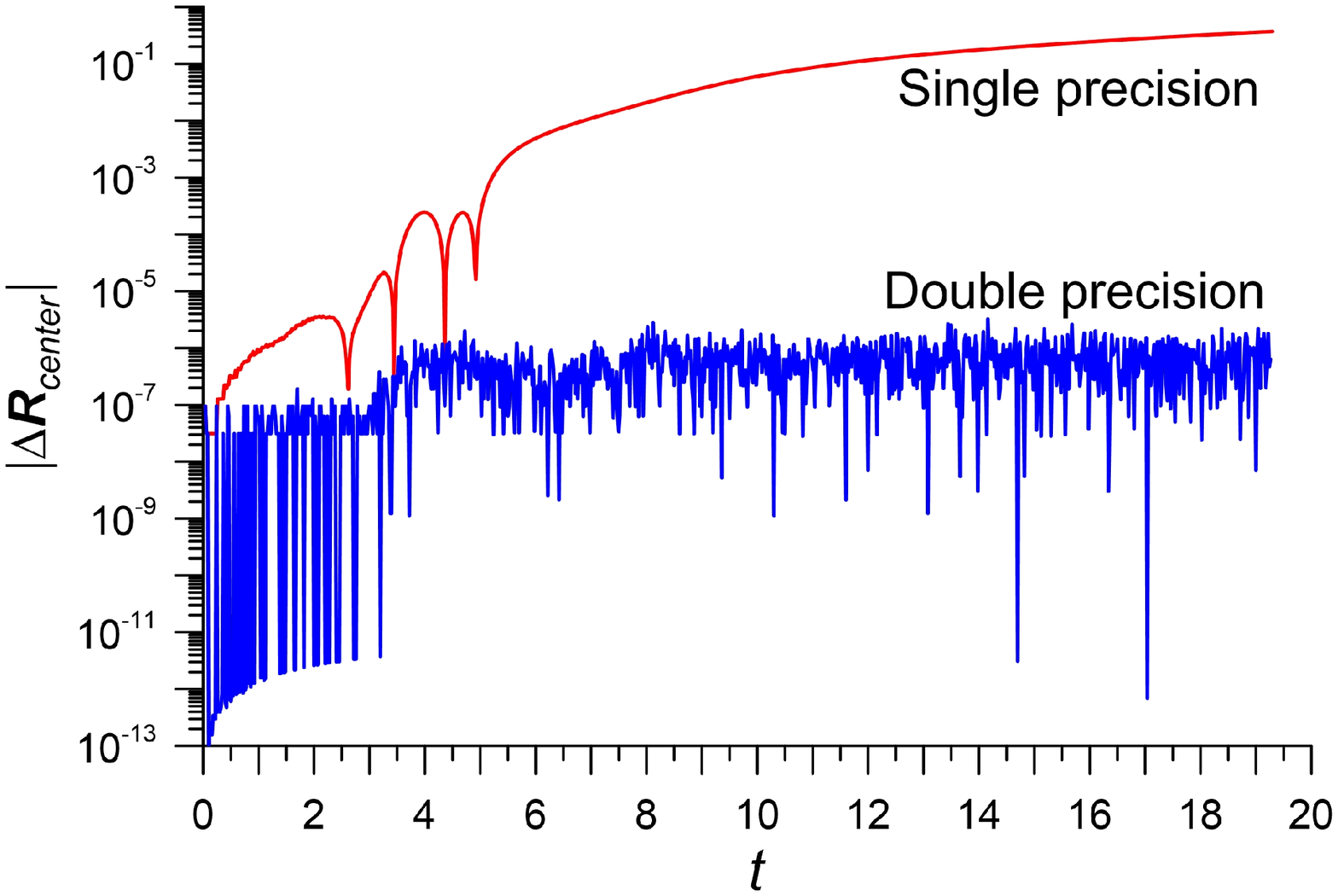}
	\vskip 0 mm
	\caption{Evolution of the absolute deviations of impulse (top panel) and mass center~(bottom panel) for the system consisting of $N = 2^{20}$ particles. Red and blue lines are single precision and double precision, respectively.}
\end{figure}

For SP all dependencies demonstrate a sharp increase of the integral values errors at $t\approx 5$~(see Figs.~5 and~6). The latter is caused by the emergence of strong asymmetric disturbances due to the development of gravitational instability in the simulated disk (Fig.~7).

As seen from Table~2 with an increase of the number of particles the conservation laws' errors for energy, momentum, and angular momentum for SP increases faster than $O(N^{1.5})$, while for DP it remains within the round-off errors.

\begin{figure}[th]
	\begin{flushright}
		\bf{Table 2}\vspace{-2mm}
	\end{flushright}
\centerline{Maximal deviation of the total impulse, total energy, and angular momentum}
\centerline{as a function of number of particles}
	\begin{center}
		\begin{tabular}{|c|c|c|c|c|c|c|}
	\hline
	$N$ & $|\Delta \mathbf{P}|_{max}$, & $|\Delta \mathbf{P}|_{max}$, & $|\Delta \mathbf{L}|_{max}/L_0$, & $|\Delta \mathbf{L}|_{max}/L_0$, & $|\Delta E|_{max}/E_0$, & $|\Delta E|_{max}/E_0$, \\
	$\times1024$  & SP        & DP                   & SP        & DP           & SP           & DP \\ \hline
	128           & 0.0040954 & $2.504\cdot10^{-15}$ & 0.0002052 & $2.621\cdot10^{-14}$ & 0.0002132 & $4.602\cdot10^{-5}$ \\ \hline
	256	          & 0.0055965 & $1.923\cdot10^{-15}$ & 0.0009059 & $3.553\cdot10^{-14}$ & 0.0005128 & $7.409\cdot10^{-5}$ \\ \hline
	512	          & 0.0138642 & $3.074\cdot10^{-15}$ & 0.0020217 & $7.017\cdot10^{-14}$ & 0.0004764 & $6.474\cdot10^{-5}$ \\ \hline
	1024          & 0.0421156 & $4.278\cdot10^{-15}$ & 0.0078611 & $7.327\cdot10^{-14}$ & 0.0029361 & $5.627\cdot10^{-5}$ \\ \hline
\end{tabular}
	\end{center}
\end{figure}

\begin{figure}[!h]
	\centering \includegraphics[width=0.99\hsize]{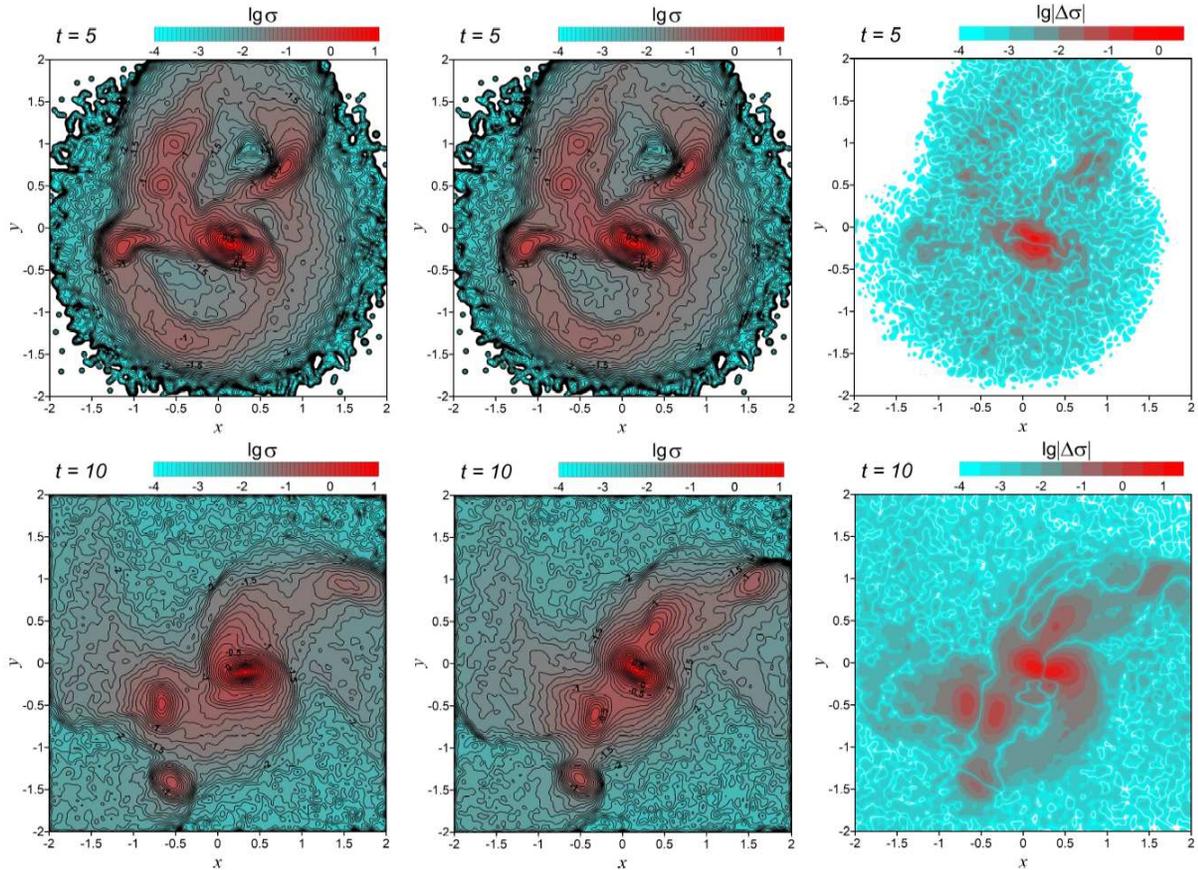}
	\vskip 0 mm
	\caption{Distribution of the surface density in the stellar disk ($N = 2^{20}$) at different times for a single precision~(left panels, $\sigma_{SP}$), double precision~(center panels, $\sigma_{DP}$) and $|\Delta \sigma| = |\sigma_{DP}-\sigma_{SP}|$ (right panels).}
\end{figure}

In Fig. 7 we show the impact of the total energy, momentum, and angular momentum conservation on the
evolution of the stellar galactic disk. There are no qualitative differences in the disk surface density distributions for SP and DP until $t\approx 5$. However, at larger times $(t > 7)$, we obtain significant quantitative and qualitative distortions of the simulation result for SP in comparison to more sophisticated DP based simulation.

\section*{Conclusion}
\hspace{0.7 cm}

In this work, we analyze the $N$-body simulations on GPUs by using a direct method of the gravitational forces calculation~(Particle-Particle algorithm) and parallel OpenMP-CUDA technologies. We found that a single-precision numbers in the second-order accuracy schemes can lead to significant quantitative and qualitative distortions of the $N$-body system evolution. We claim that this is due to significant violation of the laws of conservation of momentum, angular momentum, and energy at modeling times exceeding $10^4$  time steps of integration, which, in turn, is the result of the accumulation of round-off errors in the calculation of gravitational forces. This effect is mostly pronounced for non-axisymmetric $N$-body systems in the absence of external steady gravitational fields.


\section*{Acknowledgments}
\hspace{0.7 cm} SSK is thankful to the RFBR (grants 16-07-01037 and 16-02-00649). SAK gratefully acknowledges funding from the Russian Foundation for Basic Research (16-32-60043).
AVK is thankful to the Ministry of Education and Science of the Russian Federation (government task No.~2.852.2017/4.6). Authors also wish to thank Yulia Venichenko for useful comments which helped improve the paper.

\begin{biblio_lat}

\bibitem{bib1}
Fridman A. M., Khoperskov A. V. Physics of Galactic Disks.
\textit{Cambridge International Science Publishing Ltd}, 2013 \, 754 p. \vspace{-2mm}

\bibitem{KennedyEtal2016}
Kennedy G.F., Meiron Y., Shukirgaliyev B., Panamarev T., Berczik P. et al. The DRAGON simulations: globular cluster evolution with a million stars. \textit{Monthly Notices of the Royal Astronomical Society}, 2016, vol.\,458, no.\,2, pp.~1450--1465. DOI: 10.1093/mnras/stw274\vspace{-2mm}

\bibitem{bib11}
Khrapov S., Khoperskov A. Smoothed-particle hydrodynamics models: implementation features on GPUs. \textit{Communications in Computer and Information Science}, 2017, vol.\,793, pp.\,266--277. DOI: 10.1007/978-3-319-71255-0\_21\vspace{-2mm}

\bibitem{SmirnovSotnikova2017}	
Smirnov A.A., Sotnikova N.Ya., Koshkin A.A. Simulations of slow bars in anisotropic disk systems. \textit{Astronomy Letters}, 2017, vol.~43, no.~2, pp.~61--74. DOI: 10.1134/S1063773717020062\vspace{-2mm}

\bibitem{Klypin2017}
Comparat J., Prada F., Yepes G., Klypin A. Accurate mass and velocity functions of dark matter haloes. \textit{Monthly Notices of the Royal Astronomical Society}, 2017, vol.~469, no.~4, pp.~4157--4174. DOI: 10.1093/mnras/stx1183\vspace{-2mm}

\bibitem{KnebeEtal2018}
Knebe A., Stoppacher D., Prada F., Behrens C., Benson A. et al. MULTIDARK-GALAXIES: data release and first results. \textit{Monthly Notices of the Royal Astronomical Society}, 2018, vol.~474, no.~4, pp.~5206--5231. DOI: 10.1093/mnras/stx2662\vspace{-2mm}

\bibitem{bib2}
Hwang J.-S., Park C. Effects of hot halo gas on star formation and mass transfer during distant galaxy-galaxy encounters.
\textit{The Astrophysical Journal}, 2015, vol.\,805, pp.\,131--149. DOI: 10.1088/0004-637X/805/2/131\vspace{-2mm}

\bibitem{PortaluriDebattista2017}
Portaluri E., Debattista V., Fabricius M., Cole D.R., Corsini E. et al. The kinematics of $\sigma$-drop bulges from spectral synthesis modelling of a hydrodynamical simulation. \textit{Monthly Notices of the Royal Astronomical Society}, vol.~467, no.~1, pp.~1008--1015. DOI: 10.1093/mnras/stx172\vspace{-2mm}

\bibitem{bib10}
Khoperskov A.V., Just A., Korchagin V.I., Jalali M.A. High resolution simulations of unstable modes in a collisionless disc.
\textit{Astronomy and Astrophysics}, 2007, vol.\,473, pp.\,31--40. DOI: 10.1051/0004-6361:20066512\vspace{-2mm}

\bibitem{bib3}
Gelato S., Chernoff D.F., Wasserman I. An adaptive hierarchical particle-mesh code with isolated boundary conditions.
\textit{The Astrophysical Journal}, 1997, vol.\,480, pp.\,115--131. DOI: 10.1086/303949\vspace{-2mm}

\bibitem{bib4}
Barnes J., Hut P. A Hierarchical $O(N\log N)$ Force-Calculation Algorithm. \textit{Nature}, 1986, vol.\,324, pp.\,446--449. DOI: 10.1038/324446a0\vspace{-2mm}

\bibitem{bib5}
Greengard L. The numerical solution of the N-body problem. \textit{Computers in physics}, 1990, vol.\,4, pp.\,142--152.\vspace{-2mm}

\bibitem{Huang2016}
Huang S.-Y., Spurzem R., Berczik P. Performance analysis of parallel gravitational N-body codes on large GPU clusters. \textit{Research in Astronomy and Astrophysics}, 2016, vol.~16, no.~1, article id.~11. DOI: 10.1088/1674-4527/16/1/011\vspace{-2mm}

\bibitem{Steinberg2017}
 Steinberg O.B. Circular shift of loop body --- programme transformation, promoting parallelism. \textit{Bulletin of the South Ural State University Series: Mathematical Modelling, Programming \& Computer Software}, 2017, vol.\,10, no.~3, pp.~120--132. DOI: 10.14529/mmp170310\vspace{-2mm}

\bibitem{bib7}
Khoperskov A., Bizyaev D., Tiurina N., and Butenko M. Numerical modelling of the vertical structure and dark halo parameters in disc galaxies.
\textit{Astronomische Nachrichten}, 2010, vol.\,331, pp.\,731--745. DOI: 10.1002/asna.200911402\vspace{-2mm}

\bibitem{bib8}
Khoperskov A.V., Khoperskov S.A., Zasov A.V., Bizyaev D.V., Khrapov S.S. Interaction between collisionless galactic discs and nonaxissymmetric dark matter haloes. \textit{Monthly Notices of the Royal Astronomical Society}, 2013, vol.\,431, pp.\,1230--1239. DOI: 10.1093/mnras/stt245\vspace{-2mm}

\bibitem{bib9}
Khoperskov S.A., Vasiliev E. O., Khoperskov A. V., Lubimov V. N. Numerical code for multi-component galaxies: from N-body to chemistry and magnetic fields. \textit{Journal of Physics: Conference Series}, 2014, vol.\,510, pp.\,1--13. DOI: 10.1088/1742-6596/510/1/012011\vspace{-2mm}

\bibitem{RodionovAthanassoula2009}
 Rodionov S.A., Athanassoula E., Sotnikova N.Ya. An iterative method for constructing equilibrium phase models of stellar systems. \textit{Monthly Notices of the Royal Astronomical Society}, 2009, vol.~392, no.~2, pp.~904--916. DOI: 10.1111/j.1365-2966.2008.14110.x\vspace{-2mm}

\bibitem{bibl12}
Robert G. Bellemana, Jeroen B., Simon F., Portegies Z. High performance direct gravitational N-body simulations on graphics processing units II: An implementation in CUDA. \textit{New Astronomy}, 2008, vol.\,13, pp.\,103--112. DOI: 10.1016/j.newast.2007.07.004\vspace{-2mm}

\bibitem{bib6}
Griv E., Wang H.-H. Density wave formation in differentially rotating disk galaxies: Hydrodynamic simulation of the linear regime \textit{New Astronomy}, 2014, vol.\,30, pp.\,8--27. DOI: 10.1016/j.newast.2014.01.001\vspace{-2mm}

\bibitem{romeo2013}
Romeo A., Falstad N. A simple and accurate approximation for the Q stability parameter in multicomponent and realistically thick discs. \textit{Monthly Notices of the Royal Astronomical Society}, 2013, vol.~433, no.~2, pp.~1389-1397. DOI: 10.1093/mnras/stt809\vspace{-2mm}

\end{biblio_lat}

\end{document}